\documentclass[prl,twocolumn,showpacs,superscriptaddress,amsmath,amssymb]{revtex4}
\usepackage[utf8]{inputenc}
\usepackage{amsthm}
\usepackage[author={Ariel}]{pdfcomment}
\usepackage{qcircuit}
\usepackage{epsfig}
\usepackage{booktabs,multirow}
\newcommand{\Tr}{\mathrm{Tr}}
\newcommand{\bra}[1]{{\langle{#1}|}}
\newcommand{\ket}[1]{{|{#1}\rangle}}

\usepackage[caption=false]{subfig}
\usepackage{graphicx,bm}
\usepackage{verbatim}

\begin{document}
	\title{Selective quantum state tomography for continuous-variable systems}
	\author{Virginia Feldman}
	\affiliation{Instituto de Física, Facultad de Ingeniería, Universidad de la República, J.  Herrera  y  Reissig  565, Montevideo, Uruguay}
	\author{Ariel Bendersky}
	\affiliation{CONICET-UBA, Instituto de Investigación en Ciencias de la Computación (ICC), 1428 Buenos Aires, Argentina}
	\affiliation{Departamento de Computación, Facultad de Ciencias Exactas y Naturales, Universidad de Buenos Aires, 1428 Buenos Aires, Argentina}
	
	\begin{abstract}
		We present a protocol that allows the estimation of any density matrix element for continuous-variable quantum states, without resorting to the complete reconstruction of the full density matrix. The algorithm adaptatively discretizes the state and then, by resorting to controlled squeezing and translation operations, which are the main requirements for this algorithm, measures the density matrix element value.
		Furthermore, we show how this method can be used to achieve full quantum state tomography for continuous-variable quantum systems, alongside numerical simulations.
		
	\end{abstract}

	\maketitle

	\section{Introduction}
	
	Quantum information processing tasks always rely on preparing, manipulating, and measuring states. In order to perform such tasks, it is essential to have a tool to characterize quantum states. The protocols to characterize quantum states are usually referred to as \emph{quantum state tomography} (QST)\cite{nielsen2002,PhysRevA.87.012122,paris2004,gross2010,cramer2010,miquel2002,lvovsky2009}.
	
	When it comes to one-dimensional continuous-variable systems, the state density matrix in position representation $\rho(x,x')$ contains all the information on the state of the system.  In this article we will focus on measuring $\rho(x,x')$ for any given $x$ and $x'$. The proposed protocol is selective since it allows one to choose any $x$ and $x'$ and then estimate the associated density matrix element.
	
	Our protocol builds upon \cite{PhysRevA.87.012122} and resorts to controlled translation and squeezing operations in order to make it suitable for continuous-variable systems. Even though currently the implementation of a controlled squeezing gate is not easy, there are proposals for such a task, as shown in \cite{drechsler2020}.
	
    This article is organized as follows. First we review the protocol from \cite{PhysRevA.87.012122}. Then we present our protocol for selective measurement of continuous-variable density matrices. Finally, we show some numerical simulations of how our protocol behaves on different states, and how it can be used for full state tomography.

	\section{Quantum state tomography in finite dimension}
	
	We will now review the quantum state tomography protocol for finite dimensional systems from \cite{PhysRevA.87.012122}, upon which we build the continuous-variable one.
	
	Let $\mathcal H$ be the Hilbert space for the system we want to analyze, and $D$ its dimension. And let us consider an orthonormal basis $\mathcal B = \left\lbrace \ket{\psi_a}, a=1,...,D \right\rbrace$ for $\mathcal H$. Then, the state $\rho$ can be written as
	
	\begin{equation}
		\rho = \sum_{a, b = 1}^D \alpha_{ab} \ket{\psi_a}\bra{\psi_b}
	\end{equation}
	
	The protocol for selective and efficient quantum state tomography will be able, then, to measure any given coefficient $\alpha_{ab}$ efficiently (i.e., with resources that scale polynomially with $\log{D}$).	

	\begin{figure}[ht]
	\begin{center}
		\epsfig{file=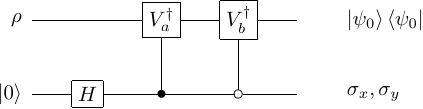, angle=0, width=0.45\textwidth}
	\end{center}
	\caption{Quantum circuit for selective and efficient quantum state tomography.\label{figTomEstNoDiagRed}}
    \label{fig_tomo}
	\end{figure}

	Consider the circuit from Fig. \ref{fig_tomo}, where the operators $V_k$ are preparation operators for the basis $\mathcal B$ such that $V_k \ket{\psi_0} = \ket{\psi_k}$. It is straightforward to see that 
	
	\begin{eqnarray}
		\Tr\left(\rho_{out} \ket{\psi_0}\bra{\psi_0}\otimes\sigma_x \right)&=&\Re{\alpha_{ab}}\\
		\Tr\left(\rho_{out} \ket{\psi_0}\bra{\psi_0}\otimes\sigma_y \right)&=&\Im{\alpha_{ab}}
	\end{eqnarray}
	where $\rho_{out}$ is the composite output state for the system and auxiliary qubit. This means that, by measuring $\sigma_x$ conditioned to having $\ket{\psi_0}$ on the first register, one has the real part of $\alpha_{ab}$, and by doing the same but with $\sigma_y$ one gets the imaginary part.

	Finally, it is shown in \cite{PhysRevA.87.012122} that the number $M$ of experimental runs in order to have an uncertainty $\epsilon$ with a probability $p$ of success is bounded, using a Chernoff bound \cite{Chernoff}, by
	\begin{equation}
		M\geq \frac{2 \ln\left(\frac{2}{p}\right) }{\epsilon^2}
		\label{efficiency}
	\end{equation}
	proving the efficiency of the method.

	\section{Selective quantum state tomography for continuous variable}
	
    As an extension of the protocol just reviewed, in this section we introduce a new protocol for the case of continuous variables, that allows to selectively measure an estimate of the density matrix element $\rho(x,x')$, for any $x$ and $x'$.
	
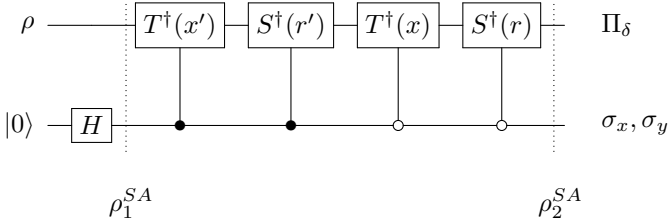
\begin{figure}[h]
        \normalsize
        \hspace{-0.5cm}
        \Qcircuit @C=0.9em @R=2.3em {
        \lstick{\rho} & \qw \ar@{.}[]+<1.3em,1em>;[d]+<1.3em,-1em> & \gate{T^{\dagger}(x')} & \gate{S^{\dagger}(r')} & \gate{T^{\dagger}(x)} & \gate{S^{\dagger}(r)} & \qw  \ar@{.}[]+<-0.4em,1em>;[d]+<-0.4em,-1em>& \rstick{\Pi_{\delta}} \\
        \lstick{\ket{0}} & \gate{H} & \ctrl{-1} & \ctrl{-1} & \ctrlo{-1} & \ctrlo{-1} & \qw & \rstick{\sigma_x,\sigma_y}\\
        &{\hspace{3em} \rho_1^{SA} \hspace{0em}}&&&&&{\hspace{1.7em} \rho_2^{SA} \hspace{2em}}&}\\
    
    \vspace{0.3cm}      
    \caption{Quantum circuit for continuous variable selective quantum state tomography.}
    \label{circuitCV}
    \end{figure}

	The proposed quantum circuit makes use of a control state and relies on controlled and anticontrolled translation $T$ and squeezing $S$ gates, as depicted in Fig. \ref{circuitCV}.
	 
    The translation operator $T(x)$ acts on a position eigenstate $\ket{y}$ as    
    \begin{equation}
        T(x)\ket{y}=\ket{x+y}.
        \label{accionT}
    \end{equation}

    The squeezing operator \cite{zhao2020high,miyata2014,yoshikawa2007} is represented by $S(r)$, with $r$ a real parameter. The action of this operator on a position eigenstate $\ket{y}$ gives a state proportional to the position eigenstate $\ket{e^{-r}y}$ \cite{kok2010introduction}. Since $S$ is unitary:
    \begin{equation}
        S(r)\ket{y}=e^{-r/2}\ket{e^{-r}y}.
        \label{accionS}
    \end{equation}
    
    The input state (system and control state) to the circuit is $\rho_0^{SA}=\rho\otimes\ket{0}\bra{0}$. A Hadamard gate is applied on the control state, yielding 
    \begin{equation}
        \rho_1^{SA}=\frac{1}{2}\rho\otimes\left(\ket{0}\bra{0}+\ket{0}\bra{1}+\ket{1}\bra{0}+\ket{1}\bra{1}\right).
    \end{equation}

    A controlled operation $CU$ acts on the combined state $\ket{\psi}\ket{0}$ as $CU( \ket{\psi}\ket{0})=\ket{\psi}\ket{0}$ and on $\ket{\psi}\ket{1}$ as $CU (\ket{\psi}\ket{1})=U\ket{\psi}\ket{1}$, while for the anticontrolled operation, $U$ only acts on $\ket{\psi}$ when the control state is $\ket{0}$. Therefore, for our circuit \ref{circuitCV}, the controlled operations $T^\dagger(x')$ and $S^\dagger(r')$ act when the control state is $\ket{1}$, while the anticontrolled gates $T^\dagger(x)$ and $S^\dagger(r)$ are applied when the state is $\ket{0}$.
    After the action of the controlled and anticontrolled gates, the combined state is given by
    \begin{equation}
    \begin{split}
        \rho_2^{SA}=&\frac{1}{2}(S^{\dagger}(r)T^{\dagger}(x)\rho T(x)S(r)\otimes \ket{0}\bra{0}\\
        & + S^{\dagger}(r')T^{\dagger}(x')\rho T(x')S(r')\otimes \ket{1}\bra{1}\\
        & + S^{\dagger}(r)T^{\dagger}(x)\rho T(x')S(r')\otimes \ket{0}\bra{1}\\
        & + S^{\dagger}(r')T^{\dagger}(x')\rho T(x)S(r)\otimes \ket{1}\bra{0}).
    \end{split}
    \end{equation}

	Taking into account that for continuous variable systems a projection measurement cannot be carried out with infinite precision, we define the projection operator around the origin as
	\begin{equation}
        \Pi_{\delta}=\int\limits_{-\delta/2}^{\delta/2}\ket{y}\bra{y}dy,
	\end{equation}
    where the  parameter $\delta$ is set by the measurement device.

	The measured average value of the operator $\Pi_\delta\otimes\sigma_x$ is
    \begin{equation}
    \begin{split}
        \textrm{Tr}&(\rho_2^{SA}\int\limits_{-\delta/2}^{\delta/2} \ket{y}\bra{y} dy\otimes \sigma_x) \\
        & = \frac{1}{2}\int\limits_{-\delta/2}^{\delta/2}\bra{y}S^{\dagger}(r)T^{\dagger}(x) \rho T(x')S(r')\ket{y}dy \\ 
        &+\frac{1}{2}\int\limits_{-\delta/2}^{\delta/2}\bra{y}S^{\dagger}(r')T^{\dagger}(x') \rho T(x)S(r)\ket{y}dy \\
        &=\textrm{e}^{-\frac{1}{2}(r+r')}\Re{\left[\int\limits_{-\delta/2}^{\delta/2}\rho(e^{-r}y+x,e^{-r'}y+x')dy\right]}.
    \end{split}
    \label{MeasureRealPart1}
    \end{equation}
    
    Considering the action of the translation and squeezing operators on a position eigenstate, that is \eqref{accionT} and \eqref{accionS}, we obtain
    \begin{equation}
    \begin{split}
     &\bra{y}S^{\dagger}(r)T^{\dagger}(x) \rho T(x')S(r')\ket{y}\\
     &=\bra{e^{-r}y}T^{\dagger}(x) \rho T(x')\ket{e^{-r'}y}=\bra{e^{-r}y+x}\rho\ket{e^{-r'}y+x'}
    \end{split}
    \end{equation}
    
    and
    \begin{equation}
    \bra{y}S^{\dagger}(r')T^{\dagger}(x') \rho T(x)S(r)\ket{y}=\bra{e^{-r'}y+x'}\rho\ket{e^{-r}y+x}
    \end{equation}

 Therefore \eqref{MeasureRealPart1} gives
 \begin{equation}
    \begin{split}
        \textrm{Tr}&(\rho_2^{SA}\int\limits_{-\delta/2}^{\delta/2} \ket{y}\bra{y} dy\otimes \sigma_x) \\
        &=\textrm{e}^{-\frac{1}{2}(r+r')}\Re{\left[\int\limits_{-\delta/2}^{\delta/2}\rho(e^{-r}y+x,e^{-r'}y+x')dy\right]}.
    \end{split}
    \label{MeasureRealPart}
    \end{equation}

\begin{figure}
\subfloat[\label{RegionFig1}]{%
  \includegraphics[width=0.8\linewidth]{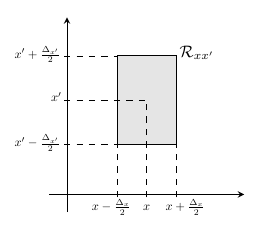}  
}\hfill
\subfloat[\label{RegionesFig2}]{
\includegraphics[width=0.8\linewidth]{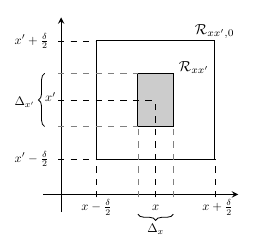}  
}
\caption{a) Region $\mathcal{R}_{xx'}$ \eqref{Region} over which the density matrix element $\rho(x,x')$ is estimated. The squeezing parameters $r$ and $r'$ allow one to select $\mathcal{R}_{xx'}$ and it is assumed that the density matrix elements do not exhibit meaningful fluctuations over this region. b) The region $\mathcal{R}_{xx',0}$, defined setting $r=r'=0$, is subdivided into smaller regions until conditions \eqref{Con1} and \eqref{Con2} are attained and the region $\mathcal{R}_{xx'}$ determined.}
\label{}
\end{figure}

    The squeezing parameters $r$ and $r'$ allow one to select a rectangular region 
    \begin{equation}
        \mathcal{R}_{xx'}=[x-\Delta_x/2,x+\Delta_x/2]\times[x'-\Delta_{x'}/2,x'+\Delta_{x'}/2],
    \label{Region}
    \end{equation}
    where $\Delta_x=e^{-r}\delta$ and $\Delta_{x'}=e^{-r'}\delta$ (Fig. \ref{RegionFig1}). The integration in \eqref{MeasureRealPart} is done over the diagonal of $\mathcal{R}_{xx'}$, which includes the density matrix element $\rho(x,x')$ we want to estimate.

    Assuming that the density matrix elements do not exhibit meaningful fluctuations over $\mathcal{R}_{xx'}$, $\rho(x,x')$ can be approximated as

    \begin{equation}
        \rho_{est}(x,x')=\frac{1}{\delta}\int\limits_{-\delta/2}^{\delta/2}\rho(e^{-r}y+x,e^{-r'}y+x')dy.
        \label{rhoest}
    \end{equation}

    Hence we can derive directly from our measurement an estimate for the real part of $\rho(x,x')$.

    Replacing the $\sigma_x$ measurement by $\sigma_y$, we obtain an estimate for the imaginary part of the density matrix element $\rho(x,x')$, namely
    \begin{equation}
    \begin{split}
        \textrm{Tr}&(\rho_2^{SA}\int\limits_{-\delta/2}^{\delta/2} \ket{y}\bra{y} dy\otimes \sigma_y)\\
        & = \textrm{e}^{-\frac{1}{2}(r+r')}\Im{\left[\int\limits_{-\delta/2}^{\delta/2}\rho(e^{-r}y+x,e^{-r'}y+x')dy\right]}. 
    \end{split}
    \end{equation}

    In order to select the aforementioned region $\mathcal{R}_{xx'}$, in other words select the squeezing parameters $r$ and $r'$, we define the weight $\varepsilon$ and seek to determine $\Delta_x$ and $\Delta_{x'}$ such that the condition
    \begin{equation}
        \int\limits_{x_i-\Delta_{x_i}/2}^{x_i+\Delta_{x_i}/2}\rho(y,y)dy\leq\varepsilon.
    \label{Condition1}    
    \end{equation}
    is verified for $x_i=x,x'$.
    
    The integral we bounded represents the probability of finding the system in the region $[x_i-\Delta_{x_i}/2,x_i+\Delta_{x_i}/2]$. 
    
    The above condition can be tested for $x_i=x$ by measuring $\ket{0}\bra{0}$ instead of $\sigma_x$ on the auxiliary qubit and for $x_i=x'$ by measuring  $\ket{1}\bra{1}$ on the auxiliary qubit. Thus, we choose the parameters $r$ and $r'$ in a manner such that the conditions
    \begin{equation}
        \textrm{Tr}(\rho_2^{SA}\Pi_\delta\otimes \ket{0}\bra{0})\leq\frac{\varepsilon}{2}
        \label{Con1} 
    \end{equation}
    and
    \begin{equation}
        \textrm{Tr}(\rho_2^{SA}\Pi_\delta\otimes \ket{1}\bra{1})\leq\frac{\varepsilon}{2}
        \label{Con2} 
    \end{equation}
     are verified.
    
    A possible way to proceed is to start by setting the squeezing parameters $r=r'=0$, that is we will work only with the detector precision $\delta$ to define a region $\mathcal{R}_{xx',0}$. If $\Delta_x$ ($\Delta_{x'}$) does not verify the condition \eqref{Con1} (\eqref{Con2}), then it is divided by two and the condition retested. The procedure is repeated until the aimed conditions are attained and our region $\mathcal{R}_{xx'}$ defined (Fig. \ref{RegionesFig2}). 
         
    It is worthy of note that, since $\rho$ is positive semi-definite,  the off-diagonal density matrix elements are bounded by $|\rho(x,x')|^2 \leq \rho(x,x)\rho(x'x')$, consequently the condition in \eqref{Condition1} also imposes a restriction of the form:
    \begin{equation}
     \left|\int\int_{\mathcal{R}_{xx'}}\rho(x,x')dxdx'\right|\leq\varepsilon,
    \end{equation}
     where the integral is done over the region ${\mathcal{R}_{xx'}}$.
      Although this does not mean that there could not exist meaningful fluctuations over $\mathcal{R}_{xx'}$, it does guarantee an upper bound for the weight of the cell.

  \begin{figure*}
\subfloat[ \label{figOsc_teo}]{%
  \includegraphics[width=0.48\linewidth]{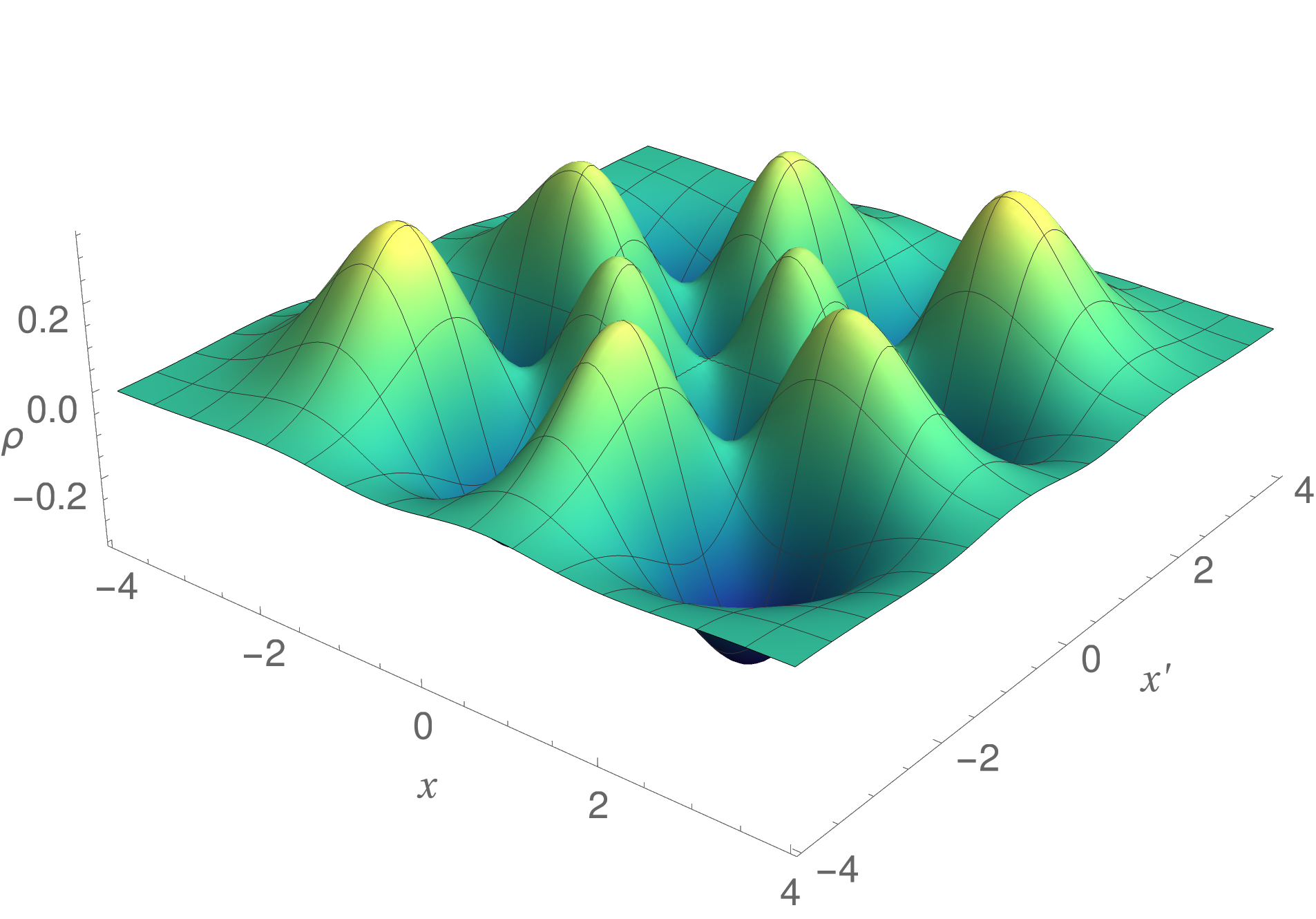}  
}\hfill
\subfloat[ \label{figOsc_sim}]{
\includegraphics[width=0.48\linewidth]{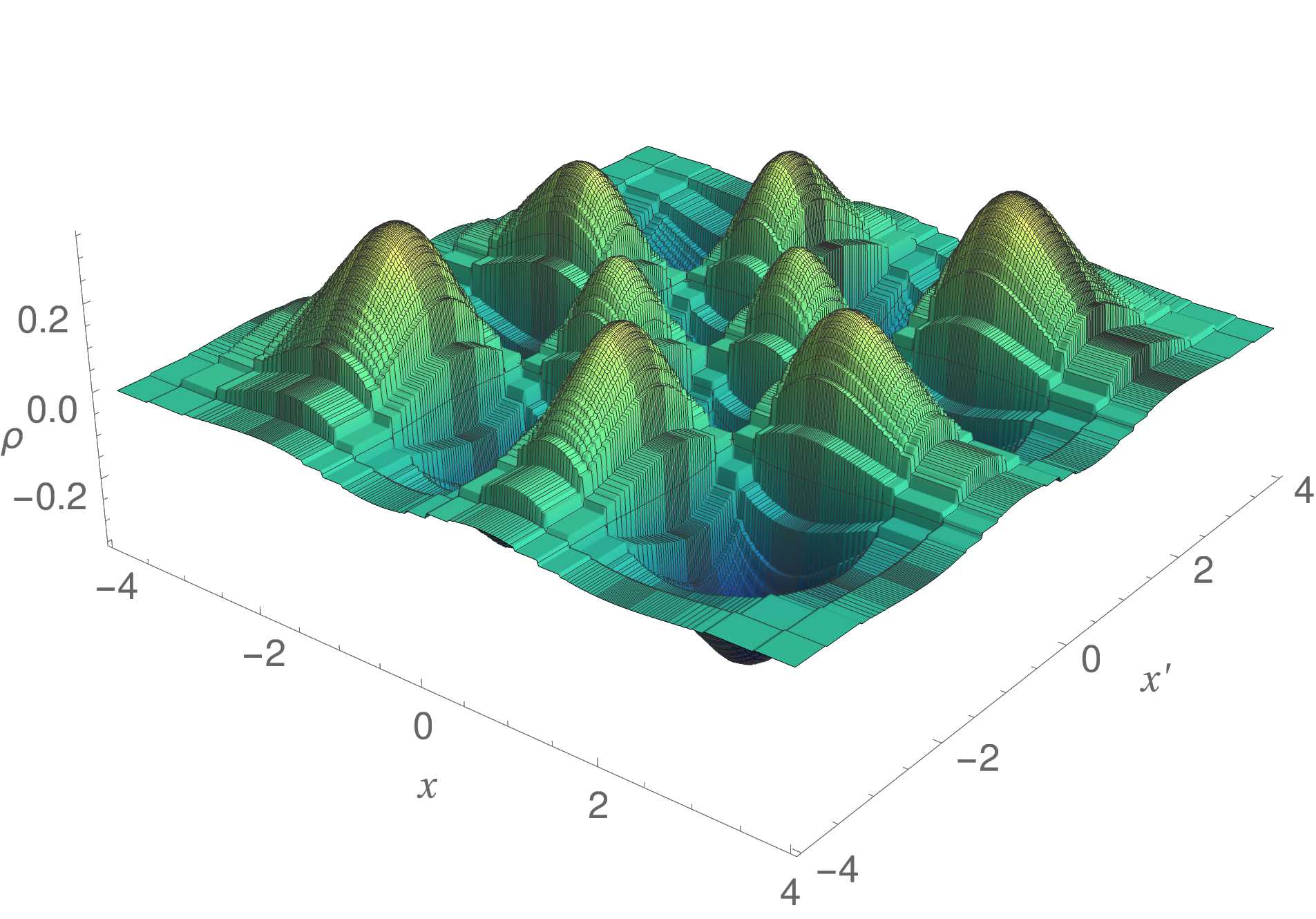}  
}
\caption{a) Theoretical $\rho(x,x')$ for the harmonic oscillator state with $n=3$. b) Reconstructed $\rho(x,x')$ for the harmonic oscillator state with $n=3$, for $\varepsilon=0.01$. For the reconstructed function, the mesh delimiting the regions $\mathcal{R}_{x_ix_j}$ in which $\rho(x,x')$ is taken as constant is shown. The detector precision was set equal to $\delta = 0.1$.}
\label{figosc}
\end{figure*}

	\begin{figure*}
\subfloat[ \label{figReSq_teo}]{%
  \includegraphics[width=0.48\linewidth]{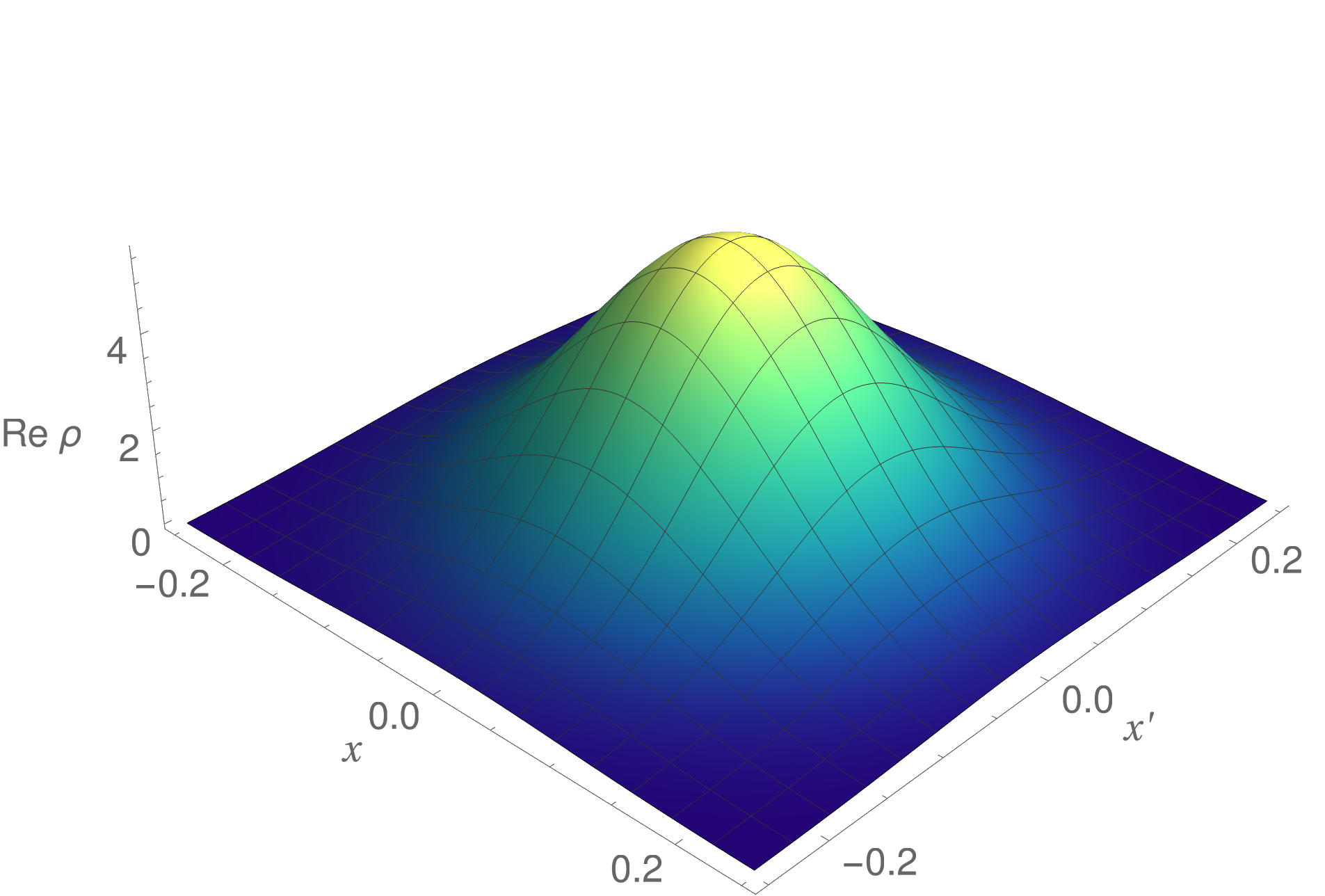}  
}\hfill
\subfloat[ \label{figReSq_sim}]{
\includegraphics[width=0.48\linewidth]{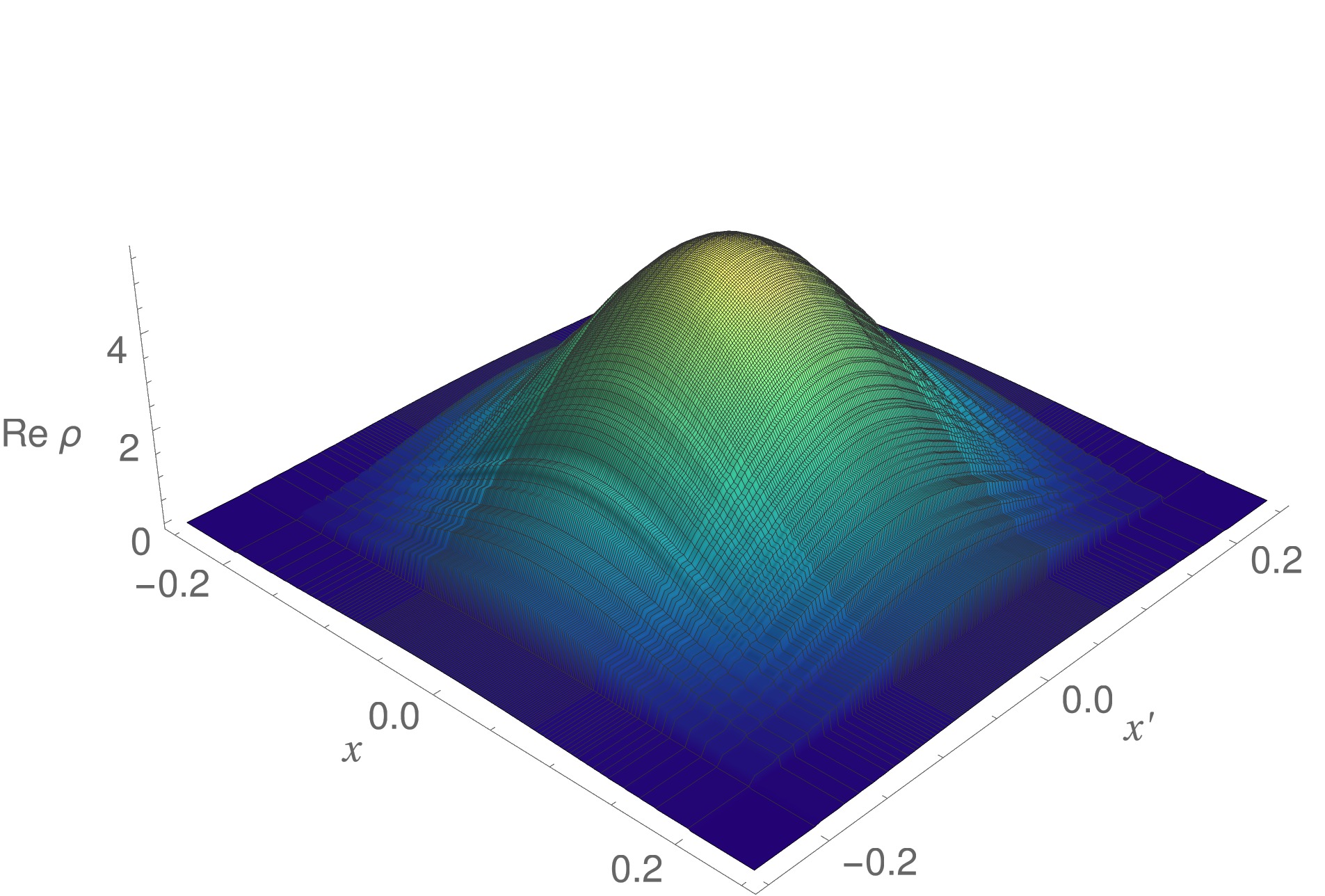}  
}\hfill
\subfloat[ \label{figImSq_teo}]{
\includegraphics[width=0.48\linewidth]{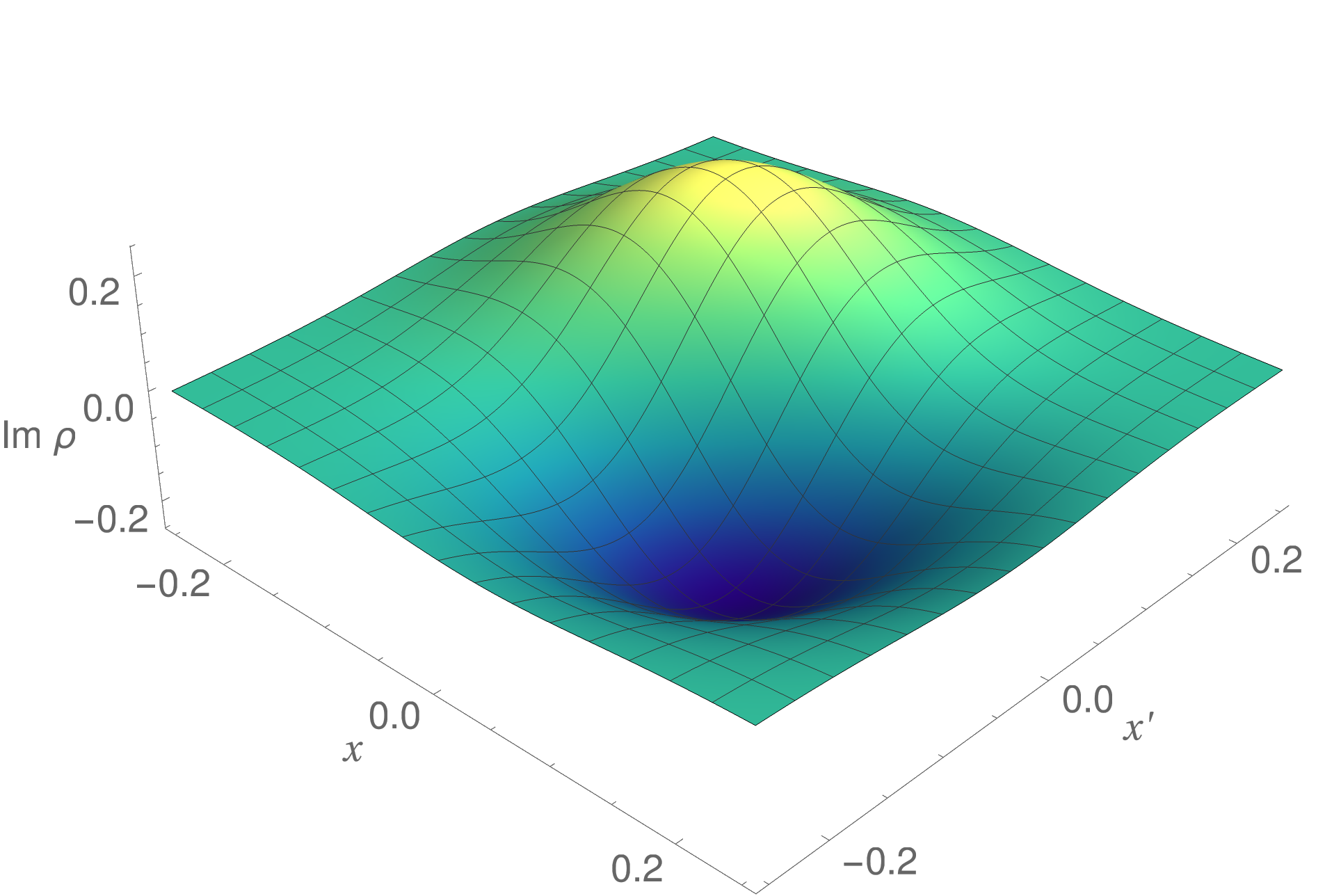}  
}\hfill
\subfloat[ \label{figReSq_sim}]{
\includegraphics[width=0.48\linewidth]{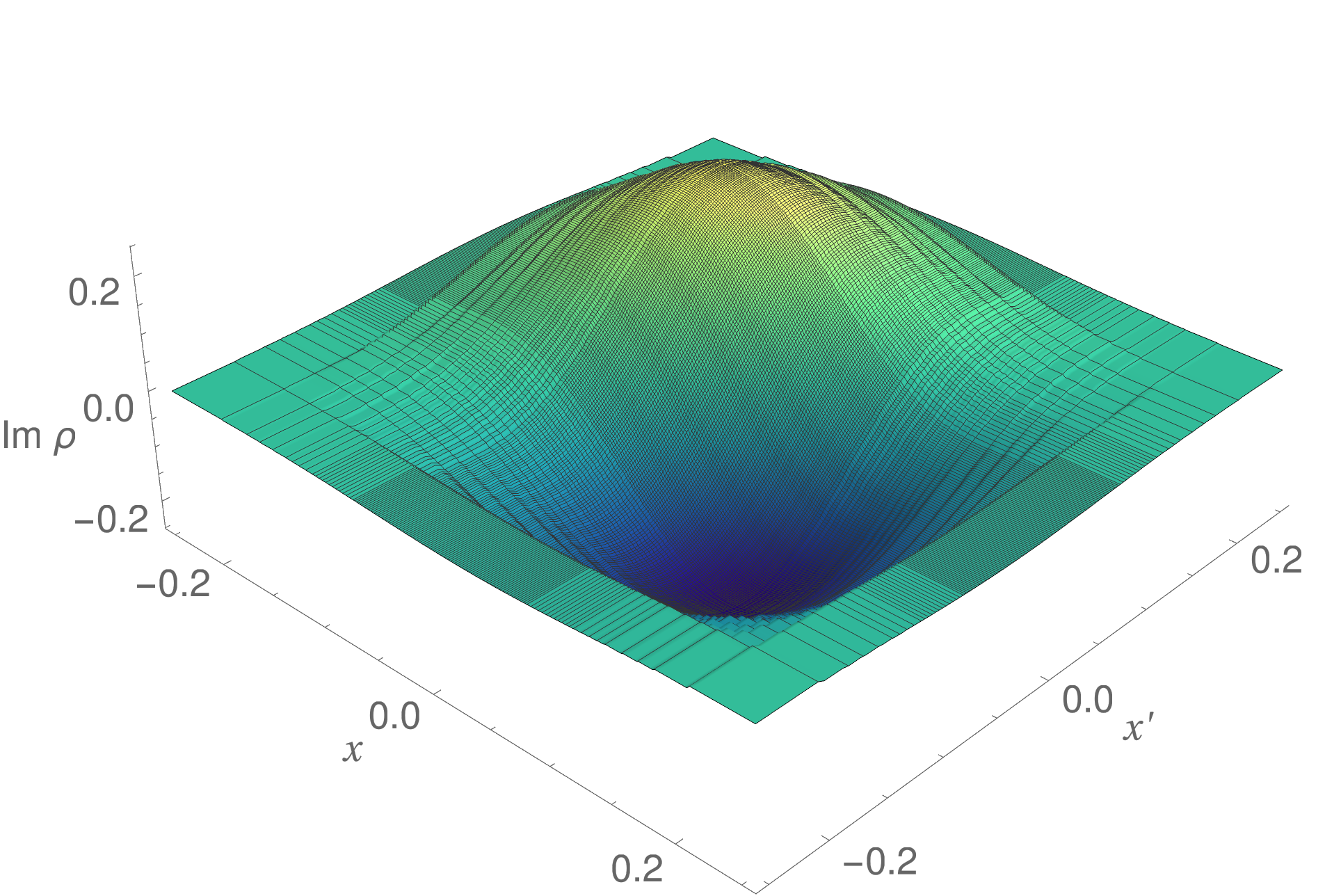}  
}
\caption{Theoretical and reconstructed $\rho(x,x')$ for a squeezed state with $\langle x\rangle=0$, $\langle p\rangle=0.5$, and $\sigma=0.1$. a) Real part of $\rho(x,x')$ b) Reconstructed real part of $\rho(x,x')$, for $\varepsilon=0.01$. c) Imaginary part of $\rho(x,x')$ d) Reconstructed imaginary part of $\rho(x,x')$, for $\varepsilon=0.01$. For the reconstructed functions, the mesh delimiting the regions $\mathcal{R}_{x_ix_j}$ in which $\rho(x,x')$ is taken as constant is shown. The detector precision was set equal to $\delta = 0.1$.}
\label{figsq}
\end{figure*}

 Each time we test condition \eqref{Con1} or condition \eqref{Con2}, the number of experimental runs $N$ required to obtain a result having an error equal or greater than $\epsilon$ with a probability $p$ or less can be determined by means of a Chernoff bound \cite{Chernoff}. Taking into consideration that each experimental run gives two possible results, corresponding to a click or no click of the detector, and that each of these results is detected at random with its corresponding probability, the number $N$ of experimental runs is bounded by
 \begin{equation}
  N\geq \frac{\ln\left(\frac{2}{p}\right) }{2\epsilon^2}.
 \end{equation}

 Once the squeezing parameters $r$ and $r'$ are set, the number of experimental runs required to obtain a resulting $\rho_{est}(x,x')$ having an error equal or greater than $\epsilon$ with a probability $p$ or less depends on the detector precision $\delta$ and the parameters $r$ and $r'$. For the estimation of $\rho_{est}(x,x')$, we need to take into consideration that each experimental run gives one of three possible results, that can be labeled as $+1$, $-1$ and $0$. The results $1$ and $0$ correspond to a click of the detector of the system state and a $\pm 1$ result for the measurement of $\sigma_x$ ($\sigma_y$) on the auxiliary one. Each of the results is detected at random with its corresponding probability. After $M$ runs of the experiment, the average values can be determined and the density matrix element $\rho(x,x')$ can be estimated by $\rho_{est}(x,x')$.  Therefore, this is a similar scenery as the one seen in \cite{PhysRevA.87.012122} and the number of experimental repetitions $M$ is once again bounded by a Chernoff bound, satisfying
    \begin{equation}
		M\geq \frac{2 \ln\left(\frac{2}{p}\right) }{\epsilon^2\delta^2e^{-(r+r')}}.
    \end{equation}

    The number of experimental repetitions scales polynomially with the uncertainty $\epsilon$.

    For this protocol to be implemented we also require an efficient implementation of the controlled squeezing and translation gates. This implementation is fundamental for quantum computation with continuous variable \cite{ menicucci2014,lloyd1999, braunstein2005,gu2009} and there had been advances towards this end in recent years, for instance \cite{drechsler2020} proposes the implementation of a controlled squeezing gate for the squeezing of trapped ions. It is also worth highlighting that displacement and squeezing gates, along with other operations, can form a universal gate set for continuous-variable quantum computation \cite{lloyd1999, hillmann2020}. 
    
    \section{Full quantum state tomography for continuous variable}
 
    The protocol introduced in the last section can also be used to achieve full QST of $\rho$. To that end, we will take a discretization approach, building the estimated function $\rho_{est}(x,x'):\mathbb{R}^2\rightarrow\mathbb{C}$ as a piecewise constant function. The regions in which the function will be considered as a constant are not fixed and will be refined based on our measurements results.
 
    In order to determine the aforementioned regions, we define the weight $\varepsilon$ and seek to obtain a partition of $\mathbb{R}$ such that any subinterval $\left[x_i-\Delta_{x_i}/2,x_i+\Delta_{x_i}/2\right]$ of the partition verifies the condition \eqref{Condition1}.
  
    The desired partition can be achieved by starting from an interval for which the integral in \eqref{Condition1} is approximately one, then dividing it into two subintervals of equal length. If a subinterval does not verify the condition, then it is likewise subdivided. The process can be repeated until the condition is attained for each subinterval.
    
    Hence, we have defined rectangular regions $\mathcal{R}_{x_ix_j}$ of the form \eqref{Region}, that is
    \begin{equation}
        \mathcal{R}_{x_ix_j}=[x_i-\Delta_{x_i}/2,x_i+\Delta_{x_i}/2]\times[x_j-\Delta_{x{_j}}/2,x_j+\Delta_{x_j}/2],
        \label{regsim}
    \end{equation}
    and we can now set 
   \begin{equation}
    \rho_{est}(x,x')=\rho_{est}(x_i,x_j),
    \label{rhosim}
    \end{equation}
    for any $(x,x')$ in $\mathcal{R}_{x_ix_j}$, with $\rho_{est}(x_i,x_j)$ as defined in \eqref{rhoest}.
    
     
     As pointed out in the previous section, this procedure guarantees an upper bound for the weight of each of the cells $\mathcal{R}_{x_ix_j}$.
    

    To test our protocol, we numerically simulated its behavior for different pure states. In order to simulate the reconstructed states we proceeded as follows.
    
    \begin{enumerate}
   
    \item An initial interval was chosen so that the bounded integral in \eqref{Condition1}  is approximately one for this interval.
    
    \item In each step, intervals which did not verify \eqref{Condition1} for our chosen weight $\varepsilon$ were subdivided into two subintervals of equal length and the condition \eqref{Condition1} retested for each new subinterval. This subdivision procedure stops when condition \eqref{Condition1} is verified  for each subinterval, thus obtaining the regions $ \mathcal{R}_{x_ix_j}$ ( \eqref{regsim}).
    
    \item The simulated reconstructed states were then obtained by setting the value of $\rho_{est}(x,x')$ constant in each region $ \mathcal{R}_{x_ix_j}$, as specified in \eqref{rhosim}.

    \end{enumerate}

    Taking into consideration that for pure states $\rho(x,x')=\psi(x)\psi(x')$, with $\psi(x)$ the pure state wave function, we studied squeezed coherent states whose wave functions can be expressed as
    \begin{equation}
        \psi_{sq}(x)=C_{sq}\textrm{exp}\left[{-\frac{(x-\langle x \rangle)^2}{2\sigma^2}+i\langle p \rangle x}\right],
        \label{sq}
    \end{equation}
    where $C_{sq}$ is a normalization constant, $\langle x \rangle$ the expected value of the position, $\langle p \rangle$ the expected value of the momentum and $\sigma^2$ the variance, which specifies the degree of squeezing, with $\sigma^2=1$ corresponding to a non squeezed state. 
    
    We also analyzed the states of the quantum harmonic oscillator with wavefunctions of the form
    \begin{equation}
        \psi_n(x)=C_ne^{-\frac{x^2}{2}}H_n(x).
        \label{osc}
    \end{equation}
    The energy level is denoted by $n$, $C_n$ is a normalization constant and $H_n(x)$ are the Hermite polynomials.
    
    The results obtained for two states of the form \eqref{osc} and \eqref{sq} are presented in Figs. \ref{figosc} and \ref{figsq} respectively.

    Even though the experimentally reconstructed states could be not physical, our results are numerically simulated, so we can use the fidelity as a measure of the distance between the state $\rho$ and our simulated estimate of it.     
    The fidelity  between the pure state $\rho=\ket{\psi}\bra{\psi}$ and our estimate  $\rho_{est}$ is given by
    
    \begin{multline}
      F(\rho,\rho_{est})= \bra{\psi}\rho_{est}\ket{\psi}=\\ =\sum\limits_{\mathcal{R}_{x_ix_j}}\rho_{est}(x_i,x_j)\iint\limits_{\mathcal{R}_{x_ix_j}}\rho(x',x)dx'dx,
    \end{multline}
    where the sum is done over all the regions $\mathcal{R}_{x_ix_j}$.
    
     The computed fidelities for different states of the form \eqref{sq} and \eqref{osc} are shown in tables \ref{FidelitySq} and \ref{FidelityOsc}.

\begin{table}[h]
\begin{center}
\setlength{\tabcolsep}{8pt} 
\renewcommand{\arraystretch}{1.5} 
\begin{tabular}{c|c|c|c|}
    \cline{2-4}
    \multirow{2}{*}{}&\multicolumn{3}{|c|}{\textbf{Fidelity}}\\
    \cline{1-4}
      \multicolumn{1}{|c|}{\boldmath{$\sigma$}}&$\varepsilon=0.01$&$\varepsilon=0.05$&$\varepsilon=0.1$\\
    \hline
    \hline
     \multicolumn{1}{|c|}{$\textbf{0.1}$} & $0.992$ & $0.987$ & $0.948$ \\
      \multicolumn{1}{|c|}{$\textbf{0.4}$} &  $0.987$ & $0.982$ & $0.945$ \\
     \multicolumn{1}{|c|}{ $\textbf{0.9}$} &  $0.981$ & $0.976$ & $0.940$\\
    \hline
    
\end{tabular}
\end{center}
\caption{Fidelity for different reconstructed squeezed quantum states as a function of $\varepsilon$. The degree of squeezing is given by the standard deviation $\sigma$, the expected value of the position is $\langle x \rangle=0$ and the expected value of the momentum is $\langle p \rangle=0.5$. The detector precision was set equal to $\delta=0.1$.}
\label{FidelitySq}
\end{table}

\begin{table}
\begin{center}
\setlength{\tabcolsep}{8pt} 
\renewcommand{\arraystretch}{1.5} 
\begin{tabular}{c|c|c|}
    \cline{2-3}
    \multirow{2}{*}{}&\multicolumn{2}{|c|}{\textbf{Fidelity}}\\
    \cline{1-3}
     \multicolumn{1}{|c|}{\textbf{n}} &$\varepsilon=0.01$&$\varepsilon=0.05$\\
    \hline
    \hline
       \multicolumn{1}{|c|}{\textbf{1}} & $0.992$ & $0.906$ \\
      \multicolumn{1}{|c|}{\textbf{2}} & $0.984$ & $0.818$ \\
      \multicolumn{1}{|c|}{\textbf{3}} & $0.981$ &  $0.953$\\
      \multicolumn{1}{|c|}{\textbf{4}} & $0.958$ & $0.846$\\
      \multicolumn{1}{|c|}{\textbf{5}} &  $0.957$ & $0.736$\\
      \multicolumn{1}{|c|}{\textbf{6}} & $0.957$ & $0.748$\\
     \multicolumn{1}{|c|}{ \textbf{7}} & $0.919$ & $0.853$\\
      \multicolumn{1}{|c|}{\textbf{8}} & $0.952$ & $0.820$\\
     \multicolumn{1}{|c|}{ \textbf{9}} & $0.944$ & $0.775$\\
     \multicolumn{1}{|c|}{ \textbf{10}} & $0.928$ & $0.743$\\
    \hline
    
\end{tabular}
\end{center}
 \caption{Fidelity for the reconstructed quantum harmonic oscillator states as a function of $\varepsilon$ for the first $10$ energy levels. The detector precision was set equal to $\delta=0.1$.}
 \label{FidelityOsc}
\end{table}

\section{Discussion}

In this paper we proposed a protocol that allows one to directly estimate any density matrix element of a quantum state for one-dimensional continuous-variable systems. This protocol is selective and works without requiring the reconstruction of the full quantum state or of a quasiprobability distribution, such as the Wigner function \cite{Wigner1932}, and without relying on inverse linear transform techniques or statistical inferences techniques for such reconstruction, procedures commonly used in the most conventional QST schemes for continuous variable systems \cite{lvovsky2009}.

We also showed how this protocol can be employed to achieve full QST. The numerical simulations for different states showed that continuous variable quantum states could be reconstructed with high fidelity, although the cases in which the density matrix elements fluctuate significantly in small regions remain a limitation found as well in other QST protocols.

\section{Acknowledgements}
This work was partially supported by Programa de Desarrollo de Ciencias B\'asicas (PEDECIBA, Uruguay), Comisi\'on Acad\'emica de Posgrado (CAP, UdelaR, Uruguay) and the ANPCyT project PICT 2018-04378. V.F. would like to thank Arturo Lezama (Facultad de Ingeniería, UdelaR) for his helpful insights.

\bibliographystyle{unsrt}
\bibliography{references}

\end{document}